\documentclass[showpacs,preprintnumbers,amsmath,amssymb,twocolumn,pra]{revtex4}
\usepackage{graphicx}
\usepackage{dcolumn}
\usepackage{bm}
\usepackage{times}
\usepackage[colorlinks,citecolor=blue,linkcolor=red]{hyperref}
\usepackage{color}
\usepackage{comment}

\begin{document}

\title{Unifying quantum heat transfer in nonequilibrium spin-boson model with full counting statistics}

\author{Chen Wang$^{1,}$}\email{wangchenyifang@gmail.com}
\author{Jie Ren$^{2,3,4,}$}\email{Xonics@tongji.edu.cn}
\author{Jianshu Cao$^{5,}$}\email{jianshu@mit.edu}

\address{
$^{1}$Department of Physics, Hangzhou Dianzi University, Hangzhou, Zhejiang 310018, P. R. China\\
$^{2}$Center for Phononics and Thermal Energy Science, School of Physics Science and Engineering,
Tongji University, 200092 Shanghai, P. R. China\\
$^{3}$China-EU Joint Center for Nanophononics, School of Physics Science and Engineering, Tongji University, 200092 Shanghai, P. R. China \\
$^{4}$Shanghai Key Laboratory of Special Artificial Microstructure Materials and Technology, School of Physics Science and Engineering, Tongji University, 200092 Shanghai, P. R. China \\
$^{5}$Department of Chemistry, Massachusetts Institute of Technology, 77 Massachusetts Avenue, Cambridge, MA 02139, USA
}

\date{\today}

\begin{abstract}
To study the full counting statistics of quantum heat transfer in a driven nonequilibrium spin-boson model, we develop a generalized nonequilibrium polaron-transformed Redfield equation with an auxiliary counting field. This
enables us to study the impact of qubit-bath coupling ranging from weak to strong regimes.
Without external modulations, we observe maximal values of both steady state heat flux and noise power at moderate coupling regimes, below which we find those two transport quantities are enhanced by the finite qubit energy bias.
With external modulations, the geometric-phase-induced heat flux shows monotonic decrease as increasing the qubit-bath coupling at zero qubit energy bias (without bias). While under finite qubit energy bias (with bias), the geometric-phase-induced heat flux exhibits an interesting reversal behavior in strong coupling regime.
Our results unify the seemingly contradictory results in weak and strong qubit-bath coupling regimes, and provide detailed dissections for the quantum fluctuation of  nonequilibrium heat transfer.
\end{abstract}

\pacs{44.90.+c, 05.60.Gg, 63.22.-m, 05.70.Ln}

%\keywords
% 44.90.+c	other topics in heat transfer
% 05.60.Gg	Quantum transport
% 44.10.+i	Heat conduction
%63.22.-m	Phonons or vibrational states in low-dimensional structures and nanoscale materials
% 05.70.Ln	Nonequilibrium and irreversible thermodynamics

\maketitle

\section{Introduction}

%1. control heat flux in nanoscale devices
Efficient realization and smart control of quantum energy transfer are of fundamental importance,
ranging from molecular electronics, quantum heat engine to quantum biology
~\cite{ydubi2011rmp,maratner2013nn,pnalbach2013pnas,MMohseni2014book,dzxu2016fp}.
In particular, the information and heat flow have been extensively studied in thermal functional devices,
spawning phononics~\cite{nbli2012rmp,jren2015aip}, where phonons are flexibly manipulated in analogy with electronic current in modern electronics~\cite{dvirasegal2005prl,dvirasegal2006prb,cwwang2006science,lwangprl2007,lfzhang2010prl,etaylor2015prl}.
In accordance with the second law of thermodynamics, it is known that the heat energy will naturally transfer from the hot source to the cold drain
driven by the thermodynamic bias (e.g., temperature), without additional external driving field.
As considering external modulations, the optimal mechanism of dynamical control can be unraveled in phononic thermal systems~\cite{dsegal2008prl,jieren2010prl,tianchen2013prb,cuchiyama2014pre}, even to pump heat against the temperature bias.

%2. prototype paradigm nesb
%a) readfield weak property
%b) nonequilibrium niba strong property
%c) exact approaches confirm
%d) polaron analytically unifies these two limiting regimes

The prototype to describe nanoscale heat transfer mediated by quantum junctions is the
nonequilibrium spin-boson model (NESB)~\cite{dvirasegal2005prl,ksaito2013prl}, which was originally proposed in the study of quantum dissipation~\cite{ajleggett1987rmp,uweiss2008book}.
The NESB is composed of a two-level system (i.e., qubit) interacting with two bosonic thermal baths under temperature bias.
Many methods have been proposed to study the microscopic mechanism of quantum heat transfer in the NESB.
Particularly, the Redfield approach has been extensively applied to analyze the weak qubit-bath coupling regime,
mainly due to the effective expression and clear physical picture~\cite{dsegal2008prl,jieren2010prl}.
The contribution of two thermal baths to the heat flux is additive,
which means that only the incoherently sequential heat-exchange processes between the qubit and baths are considered.
As such, the limitation of the Redfield approach is exposed in the strong qubit-bath coupling regime,
where the heat flux is nonlinearly dependent on the system-bath coupling strength.
In sharp contrast, the nonequilibrium nonteracting-blip approximation (NIBA) is applicable in the strong coupling limit to analytically treat
multi-phonon involved processes~\cite{dvirasegal2006prb,lnicolinjcp2011,lnicolinprb2011,dsegalpre2014},
where the nonadditive and cooperative phonon transfer processes are included.
Particularly, the appearance of turnover behavior of heat flux as a function of qubit-bath coupling strength in NESB was confirmed by NIBA, as well as by the multilayer multiconfiguration Hartree~\cite{kavelizhanin2008cpl}, quantum monte carlo schemes~\cite{ksaito2013prl} and nonequilibrium Green's function method~\cite{junjieliu2016a,junjieliu2016b,bkagarwalla1612}.
Recently, the nonequilibrium polaron transformed Redfield equation (NE-PTRE) has been proposed by the authors to analytically unify steady state heat flux in the weak and strong coupling limits, and the parity classified transfer processes are unraveled~\cite{chenwang2015sp}.

%3. from the dynamical control perspective,
%a)definition of heat pump, originally proposed by t. Thouless
%b) redfield
%c) niba

From the dynamical control perspective, the time-dependent modulation of heat transfer in NESB
has also attracted tremendous attention, enriching the transfer mechanisms~\cite{dsegal2008prl,jieren2010prl,tianchen2013prb,sgasparinetti2014njp,cuchiyama2014pre,gguarnieri2016pra,jcerrillo2016arxiv,mcarrega2016prl,lferialdi1609}.
The typical realization of the dynamical modulation is the adiabatic quantum  pump, which was originally proposed by D. J. Thouless to study
the effect of Berry phase induced quantization on the closed system transport~\cite{djthouless1983prb}.
In analogy, as the NESB is adiabatically and periodically driven by control parameters (e.g., bath temperatures),
a geometric-phase induced heat flow will contribute to the heat transfer~\cite{jieren2010prl,tianchen2013prb}.
%{\color{red}
However, previous research unraveled the seemingly contradictory results that:
in the weak qubit-bath coupling limit, the geometric-phase-induced heat flux keeps finite, independent of qubit-bath coupling strength at unbiased condition~\cite{jieren2010prl};
%}
whereas the counterpart in the strong coupling limit becomes strictly zero~\cite{tianchen2013prb}.
The statement of seemingly contradiction in below has the same meaning as expressed herein by default.
Thus, natural questions are raised: What happened in the mediate qubit-bath coupling regime? Can we propose a theory to unify the geometric-phase induced heat flux in the weak and strong coupling limits?
%More specifically, can we extend the NE-PTRE to properly analyze the adiabatic modulation of the NESB?

In the present paper, by including the full counting statistics, we introduce a generalized NE-PTRE to analyze the geometric-phase induced heat flux in the NESB.
Our NE-PTRE is able to accommodate both the sequential transfer picture in the weak coupling limit, and the multi-phonon involved nonlinear collective transfer picture in the strong coupling regime.
%It should be noted that the polaron transformed Redfield method was originally developed to study the quantum dissipative dynamics~\cite{rjsilbey1984jcp,sjang2008jcp,anazir2009prl,cklee2015jcp}.
Geometric heat pump is investigated at both unbiased and biased conditions, and the seemingly contradictory results in weak and strong coupling limits are clearly unified.
Moreover, the effect of the qubit energy bias on the geometric heat pump is analyzed in typical system-bath coupling regimes.
This work is organized as follows:
in section II, we firstly introduce the NESB and the NE-PTRE scheme that dissect the phonon transfer details.
Then in section III, by introducing full counting statistics, we develop the generalized NE-PTRE and systematically analyze counting measurements of the NESB transport.
In section IV, we firstly investigate the steady state heat flux and noise power as functions of coupling strength and qubit energy bias.
Then, we focus on the geometric-phase induced heat flux both at unbiased and biased cases, and the comparisons with Redfield and nonequilibrium NIBA are clearly demonstrated.
The final section gives a concise summary.

\section{Nonequilibrium spin-boson system}

%%==========================================
\begin{figure}[tbp]
\begin{center}
%\vspace{-2.2cm}
\includegraphics[scale=0.3]{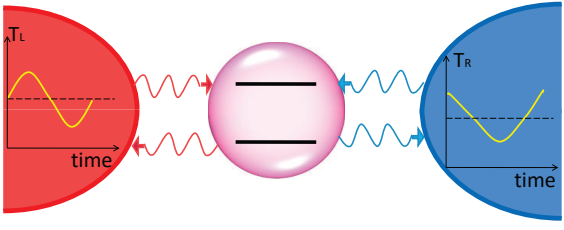}
%\vspace{-1.0cm}
\end{center}
\caption{(Color online) Schematic description of the nonequilibrium spin-boson model, composed by
central two-level qubit (purple circle) coupled to two individual thermal baths (red and blue regimes), with temperatures $T_L$ and $T_R$, respectively. The red (blue) arrowed lines describe the interaction
between the qubit and the $L$th ($R$th) bath.
For the driven nonequilibrium spin-boson model, the system parameters appear time-dependent,
e.g., $T_L(t)$ and $T_R(t)$.
}~\label{fig:fig0a}
\end{figure}
%%==========================================
\subsection{Model}
Following Ref.~\cite{chenwang2015sp}, the NESB model at Fig.~\ref{fig:fig0a}, consisting of a two-level qubit coupled to two phononic thermal baths at different temperatures
~\cite{ajleggett1987rmp,dvirasegal2005prl,jieren2010prl,uweiss2008book,ksaito2013prl}, is described as
\begin{eqnarray}~\label{h0}
\hat{H}_0=\frac{\epsilon_0}{2}\hat{\sigma}_z+\frac{\Delta}{2}\hat{\sigma}_x+\sum_{k;v=L,R}\omega_k\hat{b}^{\dag}_{k,v}\hat{b}_{k,v}  \nonumber \\
+\sum_{k;v=L,R}\hat{\sigma}_z(\lambda_{k,v}\hat{b}^{\dag}_{k,v}+\lambda^{*}_{k,v}\hat{b}_{k,v}),
\end{eqnarray}
where the qubit is specified by Pauli operators $\hat{\sigma}_z=|1{\rangle}{\langle}1|-|0{\rangle}{\langle}0|$
and $\hat{\sigma}_x=|1{\rangle}{\langle}0|+|0{\rangle}{\langle}1|$,
with $|1(0){\rangle}$ the excited (ground) state.
$\epsilon_0$ is the energy bias, and $\Delta$ is the tunneling strength between two states.
$\hat{b}^{\dag}_{k,v} (\hat{b}_{k,v})$ creates (annihilates) one phonon with energy $\omega_{k}$ and momentum $k$ in the $v$th bath, and $\lambda_{k,v}$ describes the coupling strength between the qubit and the $v$th bath.

To study the qubit-bath interaction beyond the weak coupling limit, it is helpful to transform the original Hamiltonian $\hat{H}_0$ at Eq.~(\ref{h0})
under the polaron framework by $\hat{H}=\hat{U}^{\dag}\hat{H}_0\hat{U}$~\cite{dvirasegal2006prb,anazir2009prl,tianchen2013prb},
where the unitary operator is given by $\hat{U}=e^{i\hat{\sigma}_z\hat{B}/2}$,
with the collective phononic momentum operator $\hat{B}=2i\sum_{k;v=L,R}(\frac{\lambda_{k,v}}{\omega_{k}}\hat{b}^{\dag}_{k,v}
-\frac{\lambda^{*}_{k,v}}{\omega_k}\hat{b}_{k,v})$.
Thus, the transformed Hamiltonian becomes $\hat{H}=\hat{H}_s+\hat{H}_b+\hat{V}_{sb}$.
Specifically, the re-organized two-level qubit is shown as
\begin{eqnarray}~\label{hs}
\hat{H}_s=\frac{\epsilon_0}{2}\hat{\sigma}_z+\frac{\eta\Delta}{2}\hat{\sigma}_x,
\end{eqnarray}
where the renormalization factor is given by~\cite{dvirasegal2006prb,tianchen2013prb}
\begin{eqnarray}~\label{eta}
\eta&=&{\langle}\cos\hat{B}{\rangle}\\
&=&\exp(-\sum_{v}\int^{\infty}_0d{\omega}\frac{J_v(\omega)}{\pi\omega^2}[n_v(\omega)+1/2]),\nonumber
\end{eqnarray}
with the $v$-th bath spectral function $J_v(\omega)=4\pi\sum_{k}|\lambda_{k,v}|^2\delta(\omega-\omega_k)$,
the Bose-Einstein distribution $n_v(\omega)=1/[\exp(\beta_v\omega_v)-1]$,
and inverse of the $v$th bath temperature $\beta_v=1/k_BT_v$.
The noninteracting phonon baths are characterized as $\hat{H}_b=\sum_{v=L,R}\hat{H}_v$, with
$\hat{H}_v=\sum_{k}\omega_k\hat{b}^{\dag}_{k,v}\hat{b}_{k,v}$.
The qubit-bath interaction is expressed as
\begin{eqnarray}~\label{vsb}
\hat{V}_{sb}=\frac{\Delta}{2}[(\cos\hat{B}-\eta)\hat{\sigma}_x+\sin\hat{B}\hat{\sigma}_y],
\end{eqnarray}
of which the thermal average vanishes, i.e., ${\langle}\hat{V}_{sb}{\rangle}=0$.
Hence, it may be appropriate to perturbatively obtain the equation of motion for the two-level qubit in the polaron picture.
It should be noted that in many traditional approaches including many-phonon involved processes, e.g., nonequilibrium noninteracting blip approximation (NIBA),
the system-bath interaction $\hat{V}=\frac{\Delta}{2}(\cos\hat{B}\hat{\sigma}_x+\sin\hat{B}\hat{\sigma}_y)$ is directly perturbed
~\cite{dvirasegal2006prb,tianchen2013prb}.
%rather not the re-organized term $\hat{V}_{sb}$.
However, actually $\hat{V}$ should not be treated as a perturbation due to the non-negligible contribution of
${\langle}\hat{V}{\rangle}{\neq}0$.
In contrast, $\hat{V}_{sb}=\hat{V}-{\langle}\hat{V}{\rangle}$ at Eq.~(\ref{vsb}) may be properly perturbed in accordance with the perturbation theory~\cite{chenwang2015sp}.
%regardless of the tunneling strength and qubit-bath coupling strength.

In this paper, the spectral function of phonon baths is characterized as
$J_v(\omega)=\pi\alpha_v\omega^{s}\omega^{1-s}_{c,v}e^{-\omega/\omega_{c,v}}$,
which is typically considered in the quantum transfer studies of nanojunction systems
~\cite{uweiss2008book,sjang2008jcp,anazir2009prl,cklee2015jcp,pnalbach2010njp,dzxu2016njp}.
$\alpha_{v}$ is the system-bath coupling strength in the order $\alpha_{v}{\sim}|\lambda_{k,v}|^2$,
and $\omega_{c,v}$ is the cut-off frequency of the $v$-th phonon bath.
Without loss of generality, we consider the super-Ohmic spectrum $s=3$ in this study.
Hence, the renormaliztion factor is specified as
$\eta=\exp\{-\sum_{v=L,R}\alpha_v[-1+\frac{2}{(\beta_v\omega_{c,v})^2}\psi_1(1/\beta_v\omega_{c,v})]/2\}$,
with the trigamma function $\psi_1(x)=\sum^{\infty}_{n=0}\frac{1}{(n+x)^2}$.
Moreover, in the weak coupling limit $\alpha_{v}{\ll}1$, the normalization factor $\eta$ becomes $1$.
While in the strong coupling regime $\alpha_{v}{\gg}1$, it vanishes ($\eta=0$).
\begin{comment}
It should be noted that in some heat transfer works, the Ohmic baths ($s=1$) are also considered~\cite{dvirasegal2005prl,dvirasegal2006prb},
and the nonequilibrium NIBA is considered to analyze the steady state transfer behaviors.
In the present study for Ohmic baths case, the renomalization factor $\eta$ becomes diminished, due to the infrared divergence
$\lim_{\omega{\rightarrow}0}{J_v(\omega)}/{\omega^2}{\rightarrow}\infty$, regardless of the system-bath coupling strength.
Then, $\hat{V}$ is equivalent to the re-organized interaction term $\hat{V}_{sb}$, and is able to be directly perturbed.
Accordingly, the nonequilibrium polaron transformed Redfield scheme is reduced to the nonequilibrium NIBA.
\end{comment}

%%==========================================
\begin{figure}[tbp]
\begin{center}
%\vspace{-2.2cm}
\includegraphics[scale=0.35]{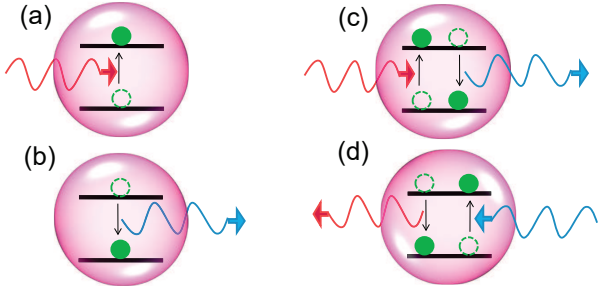}
%\vspace{-1.0cm}
\end{center}
\caption{(Color online) Representative processes involving phonons in quantum heat transfer:
(a) and (b) denote single phonon involved sequentially incoherent processes, $Q_L(\omega)$ and $Q_R(-\omega)$, respectively;
(c) and (d) depict two-phonons involved co-tunneling processes, $Q_L(\omega)Q_R(-\omega)$ and $Q_R(\omega)Q_L(-\omega)$, respectively.
}~\label{fig:fig0b}
\end{figure}
%%==========================================

\subsection{Nonequilibrium polaron transformed Redfield equation}
We note that the PTRE method was originally developed to study the quantum dissipative dynamics~\cite{rjsilbey1984jcp,sjang2008jcp,pnalbach2010njp,anazir2009prl,cklee2015jcp}, with a single bath. Here we handle a system coupled with at least two baths at nonequilibrium condition.
It is known that the re-organized system-bath interaction $\hat{V}_{sb}$ can be treated as a perturbation~\cite{chenwang2015sp}.
Based on the Born-Markov approximation and the second-order perturbation theory, we obtain the
NE-PTRE as
\begin{equation}~\label{qme1}
\frac{{\partial}\hat{\rho}}{{\partial}t}=-i[\hat{H_s},\hat{\rho}]+\sum_{l=e,o}\sum_{\omega,\omega^{\prime}=0,\pm\Lambda}
\Gamma_l(\omega)[\hat{P}_l(\omega)\hat{\rho},\hat{P}_l(\omega^{\prime})]+H.c.,
\end{equation}
where $\hat{\rho}$ is the reduced density matrix of the qubit in the polaron picture,
$\Lambda=\sqrt{\epsilon^2_0+\eta^2\Delta^2}$ is the energy gap in the eigenbasis,
%{\color{red}
and $\hat{P}_{e(o)}(\omega)$ is the eigenstate transition projector (see note.~\cite{transitionprojector}), of which the relation with Pauli matrices is given by
$\hat{\sigma}_{x(y)}(-\tau)=\sum_{\omega=0,\pm\Lambda}\hat{P}_{e(o)}(\omega)e^{i\omega\tau}$.
%}
The transition rates are
\begin{eqnarray}~\label{gamma1}
\Gamma_{o}(\omega)&=&(\frac{\eta\Delta}{2})^2\int^{\infty}_0d{\tau}e^{i\omega\tau}\sum^{\infty}_{n=0}\frac{Q(\tau)^{2n+1}}{(2n+1)!},\\
\Gamma_{e}(\omega)&=&(\frac{\eta\Delta}{2})^2\int^{\infty}_0d{\tau}e^{i\omega\tau}\sum^{\infty}_{n=1}\frac{Q(\tau)^{2n}}{(2n)!},
~\label{gamma2}
\end{eqnarray}
with the collective phonon propagator $Q(\tau)=\sum_{v=L,R}Q_v(\tau)$, and
\begin{eqnarray}~\label{qvf}
Q_v(\tau)=\int^{\infty}_0{d\omega}\frac{J_v(\omega)}{\pi\omega^2}
[n_v(\omega)e^{i\omega\tau}+(1+n_v(\omega))e^{-i\omega\tau}].
\end{eqnarray}
%describing the phonon absorption and emission processes.
From expressions of correlation functions $\Gamma_{e(0)}(\omega)$, it is clearly shown that phonon transfer processes are classified by the even and odd parity contributions.
Specifically, $\Gamma_o(\tau)$ describes the transfer processes including odd phonon numbers from two baths.
%{\color{red}
The lowest order term  $\Gamma^{(1)}_o(\omega)$ contains terms
$\frac{(\eta\Delta)^2}{8}[Q_L(\omega)+Q_R(\omega)]$, with individual bath contribution
$Q_v(\omega)=\int^{\infty}_{-\infty}{d\tau}e^{i\omega\tau}Q_v(\tau)$ at the transition energy $\omega={\pm}\Lambda$, so that the lowest odd parity exhibits sequential-tunneling behavior depicted at Figs.~\ref{fig:fig0b}(a) and \ref{fig:fig0b}(b)~\cite{dvirasegal2005prl,jieren2010prl}.
%This order demonstrates that the excitation and relaxation of phonons mediated by two-level system is driven by two baths separately, or say additively.
While $\Gamma_e(\omega)$ shows cooperative heat transfer processes involving even phonon numbers.
The corresponding lowest-order even term $\Gamma^{(1)}_e(0)$ describes the co-tunneling effect at Figs.~\ref{fig:fig0b}(c) and \ref{fig:fig0b}(d)~\cite{truokola2011prb}, which contains
$\frac{(\eta\Delta)^2}{8\pi}\int^{\infty}_{-\infty}d{\omega}Q_L(\omega)Q_R(-\omega)=\frac{(\eta\Delta)^2}{8\pi}\int^{\infty}_{0}d{\omega}[Q_L(\omega)Q_R(-\omega)+Q_R(\omega)Q_L(-\omega)]$.
%}
This demonstrates the physical picture that as the left bath releases thermal energy $\omega$,
the right bath gains the equivalent quanta simultaneously, and the two-level system only has the virtual processes of excitation and relaxation so that it keeps intact.
Apparently, these contributions from two baths are involved non-additively.
Moreover, we can obtain arbitrary order contribution to heat transfer processes systematically
by applying the Taylor expansion.

Particularly, without bias ($\epsilon_0=0$) the steady state densities can be obtained analytically in the local basis, where the diagonal and off-diagonal terms
are~\cite{chenwang2015sp}
\begin{eqnarray}~\label{pij}
P_{11}&=&P_{00}=1/2,\\
P_{10}&=&P_{01}=\frac{1}{2}\frac{\textrm{Re}[\Gamma_o(-\Lambda)]-\textrm{Re}[\Gamma_{o}(\Lambda)]}
{\textrm{Re}[\Gamma_o(-\Lambda)]+\textrm{Re}[\Gamma_o(\Lambda)]},%\nonumber
\end{eqnarray}
with the element $P_{ij}=\lim_{t{\rightarrow}\infty}{\langle}i|\hat{\rho}(t)|j{\rangle}$
($|i{\rangle}$ depicts the qubit state), energy gap $\Lambda=\eta\Delta$,
and $\textrm{Re}[\Gamma_{o(e)}(\omega)]$ the real part of $\Gamma_{o(e)}(\omega)$.

\section{Full counting statistics of NESB}

%measurement steps of the heat quantity  for initial time, and at time tau
% the change of quantity
We study the statistics of the transported heat ${\Delta}q_{\tau}=\sum_k\omega_k\Delta{n_{k,v}}$ in NESB,
from the system to the $v$-th phonon bath during a time interval $\tau$,
with $\Delta{n}_{k,v}$ the change of phonon number to initial one with momentum $k$.
The specific measurement of $\Delta{q_{\tau}}$ can be conducted as follows:
Initially at the time $t=0$, we introduce a projector $\hat{K}_{q_{0}}=|q_0{\rangle}{\langle}q_0|$ to measure the quantity $\hat{H}_v=\sum_{k}\omega_k\hat{b}^{\dag}_{k,v}\hat{b}_{k,v}$ in the $v$-th bath, giving $q_0=\sum_k\omega_kn_{k,v}(0)$.
After a finite time $\tau$ evolution of the system by coupled to thermal baths,
we again perform the projector $\hat{K}_{q_{\tau}}=|q_{\tau}{\rangle}{\langle}q_{\tau}|$
to obtain the measurement outcome $q_{\tau}=\sum_k\omega_kn_{k,v}(\tau)$.
Hence, the number difference is given by $\Delta{n}_{k,v}=n_{k,v}(\tau)-n_{k,v}(0)$.
Meanwhile, the joint probability to measure $q_0$ at $t=0$ and $q_{\tau}$ at $t=\tau$ is defined as~\cite{mesposito2009rmp}
\begin{eqnarray}
\textrm{Pr}[q_{\tau},q_0]=\textrm{Tr}_{s,b}\{\hat{K}_{q_{\tau}}e^{-i\hat{H}_0\tau}\hat{K}_{q_0}\hat{\rho}_0\hat{K}_{q_0}
e^{i\hat{H}_0\tau}\hat{K}_{q_{\tau}}\},
\end{eqnarray}
with the trace over both the qubit and thermal baths.
%connect cumulant generating function with the probability during time interval tau
Based on the joint probability $\textrm{Pr}[q_{\tau},q_0]$, we introduce the probability of measuring $\Delta{q}_{\tau}$ during the time interval $\tau$ as
\begin{eqnarray}
\textrm{Pr}_{\tau}(\Delta{q}_{\tau})=\sum_{q_{\tau},q_0}\delta(\Delta{q}_{\tau}-(q_{\tau}-q_0))\textrm{Pr}[q_{\tau},q_0].
\end{eqnarray}
Then, the cumulant generating function of the statistics can be defined as
\begin{eqnarray}
G_{\tau}(\chi)=\ln{\int{d\Delta{q}_{\tau}}\textrm{Pr}_{\tau}(\Delta{q}_{\tau})e^{i\chi\Delta{q}_{\tau}}},
\end{eqnarray}
with $\chi$ the counting field parameter.

%to calculate cumulant generating function, we develop modified NE-PTRE in the context of fsc.
%show the counting detail and obtain the modified quantum master eqaution
To quantitatively express the cumulant generating function, we introduce the NE-PTRE accompanied by the full counting statistics.
Assuming the quantum system connects to two baths (labeled by $L$ and $R$),
we measure the transported heat from the system to the $R$-th bath, in the context of the $\chi$-dependent NE-PTRE.
Then, we add the counting projector to the Hamiltonian $\hat{H}_0$ at Eq.~(\ref{h0}) to generate
$\hat{H}_{0}(\chi)=e^{i\chi\hat{H}_R/2}\hat{H}_0e^{i\chi\hat{H}_R/2}$~\cite{mesposito2009rmp,mcampisi2011rmp}, shown as
\begin{eqnarray}
\hat{H}_0(\chi)=\frac{\epsilon_0}{2}\hat{\sigma}_z+\frac{\Delta}{2}\hat{\sigma}_x+\sum_{k;v=L,R}\omega_k\hat{b}^{\dag}_{k,v}\hat{b}_{k,v}   \nonumber \\
+\sum_{k;v=L,R}\hat{\sigma}_z(e^{i\chi\omega_k\delta_{v,R}/2}\lambda_{k,v}\hat{b}^{\dag}_{k,v}+H.c.).
\end{eqnarray}
Similar to the transformation scheme in the NE-PTRE~\cite{chenwang2015sp}, we perform a generalized polaron transformation to result in $\hat{H}_{\chi}=\hat{U}^{\dag}_{\chi}\hat{H}_0(\chi)\hat{U}_{\chi}$,
with the unitary operator $U_{\chi}=e^{i\hat{\sigma}_z\hat{B}_{\chi}/2}$ and
$\chi$-dependent phonon collective momentum $\hat{B}_{\chi}=2i\sum_{k,v}(e^{i\chi\omega_k\delta_{v,R}/2}\frac{\lambda_{k,v}}{\omega_k}\hat{b}^{\dag}_{k,v}
-H.c.)$,
As such, the transformed Hamiltonian is expressed as
$\hat{H}_{\chi}=\hat{H}_s+\hat{H}_b+\hat{V}_{sb}(\chi)$.
Particularly, the re-organized qubit-bath coupling is modified with counting field, as
\begin{eqnarray}
\hat{V}_{sb}(\chi)=\frac{\Delta}{2}[(\cos\hat{B}_{\chi}-\eta)\hat{\sigma}_x+\sin\hat{B}_{\chi}\hat{\sigma}_y],
\end{eqnarray}
which includes both the information of counting measurement and the multi-phonon involved nonlinear processes.
Whereas $\hat{H}_s$ and $\hat{H}_b$ keep unchanged.
It should be noted that thermal average of the interaction term vanishes ${\langle}\hat{V}_{sb}(\chi){\rangle}=0$ due to the parity symmetry.
%{\color{red}
Moreover, the magnitude of second-order correlated contribution of $\hat{V}_{sb}(\chi)$ is quite small, compared to $\hat{H}_s$ at Eq.~(\ref{hs}).
Hence, the perturbation of $\hat{V}_{sb}(\chi)$ can be properly carried out, like the derivation of Eq.~(\ref{qme1}).
%}
Considering the Born-Markov approximation, we perturb $\hat{V}_{sb}(\chi)$ up to the second-order
and obtain the generalized NE-PTRE in the context of full counting statistics:
\begin{align}~\label{qme2}
\frac{{\partial}\hat{\rho}_{\chi}}{{\partial}t}=-i[\hat{H_s},\hat{\rho}_{\chi}]+\sum_{l=e,0}\sum_{\omega,\omega^{\prime}=0,\pm\Lambda}
[(\Gamma^{\chi}_{l,-}(\omega)+\Gamma^{\chi}_{l,+}(\omega^{\prime}))\times  \nonumber  \\
 \hat{P}_l(\omega^{\prime})\hat{\rho}_{\chi}\hat{P}_l(\omega)-(\Gamma_{l,+}(\omega)\hat{P}_l(\omega^{\prime})\hat{P}_{l}(\omega)\hat{\rho}_{\chi}+H.c.)],
\end{align}
where $\hat{\rho}_{\chi}$ is the reduced two-level system (qubit) density operator under the counting field,
$\hat{P}_l(\omega)$ is eigenstate transition projector~\cite{transitionprojector},
and energy gap is $\Lambda=\sqrt{\epsilon^2_0+\eta^2\Delta^2}$.
The transition rates are expressed as
\begin{eqnarray}
\Gamma^{\chi}_{e,\sigma}(\omega)&=&(\frac{\eta\Delta}{2})^2{\int}^{\infty}_0d{\tau}e^{i\omega\tau}[{\cosh}Q(\sigma\tau-\chi)-1],\\
\Gamma^{\chi}_{o,\sigma}(\omega)&=&(\frac{\eta\Delta}{2})^2{\int}^{\infty}_0d{\tau}e^{i\omega\tau}{\sinh}Q(\sigma\tau-\chi),
\end{eqnarray}
where the modified single phonon propagator becomes
$Q(\tau-\chi)=Q_L(\tau)+Q_R(\tau-\chi)$.
%In absence of the counting field parameter $\chi=0$, $\gamma^{\chi}_{e(0)}(\sigma\tau)$ reduces to terms
%at Eqs.~(\ref{gamma1}) and (\ref{gamma2}).

\section{Results and discussion}
In this section, we apply the generalized nonequilibrium polaron-transformed Redfield equation (NE-PTRE) with auxiliary counting field, to study the steady state heat transfer, as well as the geometric-phase induced heat transfer under adiabatic time-dependent modulations.

\subsection{Steady state heat transfer}
%general solution time-independent system, left right eigen-states
By re-arranging the NE-PTRE in the Liouville space~\cite{chenwang2015sp},
the equation of motion for the two-level qubit at Eq.~(\ref{qme2}) is expressed as
\begin{eqnarray}~\label{lde}
\frac{{\partial}}{{\partial}t}|\rho_{\chi}{\rangle}=-\mathcal{\hat{L}}_{\chi}|\rho_{\chi}{\rangle},
\end{eqnarray}
where the vector form of the density matrix is
$|\rho_{\chi}{\rangle}=[P^{\chi}_{11},P^{\chi}_{00},P^{\chi}_{10},P^{\chi}_{01}]^T$
with $P^{\chi}_{ij}={\langle}i|\hat{\rho}_{\chi}|j{\rangle}$,
and $\mathcal{\hat{L}}_{\chi}$ is the Liouvillion super-operator.
In absence of the counting field parameter ($\chi=0$), the element of density operator $P^{\chi}_{ij}$ reduces to the conventional  $P_{ij}$. % at Eq.~(\ref{pij}).
%{\color{red}
Based on the dynamical equation at Eq.~(\ref{lde}), the reduced density matrix at time $t$  is
given by $|\rho_{\chi}(t){\rangle}=\exp{(-\mathcal{\hat{L}}_{\chi}t)}|\rho_{\chi}(0){\rangle}$,
with $|\rho_{\chi}(0){\rangle}$ the initial state.
Hence, the cumulant function can be expressed as $\mathcal{Z}_{\chi}(t)={\langle}\mathrm{I}|\rho_{\chi}(t){\rangle}$~\cite{mcampisi2011rmp},
%}
with the unit vector defined as ${\langle}\mathrm{I}|=[1,1,0,0]$.
Consequently, the cumulant generating function after long-time evolution can be obtained by
${G}_t({\chi})=\frac{1}{t}\ln\mathcal{Z}_{\chi}(t)$,
and the corresponding $n$-th cumulant of heat current fluctuations can be generated as
$J^{(n)}(t)={\langle}\hat{Q}^{n}{\rangle}/t
=\frac{{\partial}^{n}{G}_t(\chi)}{{\partial}(i\chi)^{n}}|_{\chi=0}$.
When external modulation is absent, i.e., $\mathcal{L}_{\chi}$ is time-independent,
if we focus on the steady state solution,
the cumulant generating function is simplified to ${G}({\chi})=-E_0(\chi)$,
where $E_0(\chi)$ is the ground state energy of the super-operator $\mathcal{\hat{L}}_{\chi}$. The corresponding left and right eigenvectos are denoted as ${\langle}\Phi_{\chi}|$
and $|\Psi_{\chi}{\rangle}$,
which fulfill the normalization relation ${\langle}\Phi_{\chi}|\Psi_{\chi}{\rangle}=1$. %$\textrm{Tr}_s\{{\langle}\Phi_{\chi}|\Psi_{\chi}{\rangle}\}=1$.
%{\color{red}
In particular, the steady state heat flux is the first cumulant ${J}=-\frac{{\partial}E_0(\chi)}{{\partial}(i\chi)}|_{\chi=0}$,
and the noise power is the second cumulant ${J}^{(2)}=-\frac{{\partial}^2E_0(\chi)}{{\partial}(i\chi)^2}|_{\chi=0}$.
%}

\subsubsection{Unbiased condition: $\epsilon_0=0$}
\begin{figure}%[tbp]
\begin{center}
%\vspace{-2.2cm}
\includegraphics[scale=0.43]{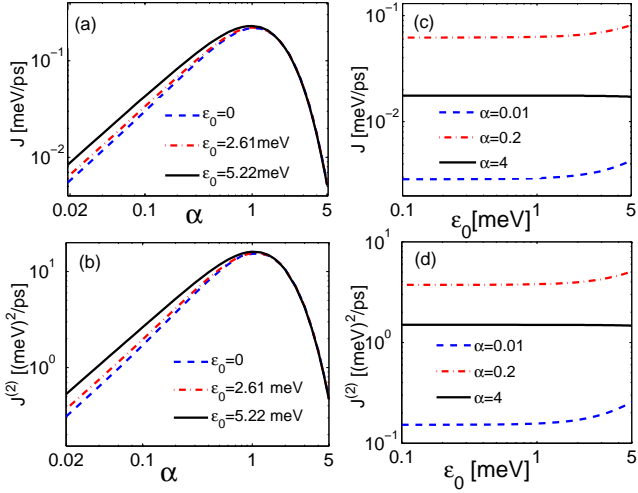}
%\vspace{-1.0cm}
\end{center}
\caption{(Color online) Behaviors of the steady state heat flux and noise power:
(a) and (b) by varying system-bath coupling strength; (c) and (d) by tuning qubit energy bias, respectively.
The other parameters are given by $\Delta=5.22$~meV, $\omega_c=26.1$~meV, $T_L=150$~K and $T_R=90$~K.
}~\label{fig:fig3}
\end{figure}

We firstly investigate the steady state heat transfer at Fig.~\ref{fig:fig3}, where the system parameters are time-independent.
without bias ($\epsilon_0=0$), the authors have shown at Ref.~\cite{chenwang2015sp} that the heat flux can be analytically solved in a wide system-bath coupling regime,
by applying the NE-PTRE.

%%==========================================
%%==========================================

%Starting from the eigen-equation
%$\mathcal{\hat{L}}_{\chi}|\Psi_{\chi}{\rangle}{\rangle}=E_0(\chi)|\Psi_{\chi}{\rangle}{\rangle}$,
%if we take the derivative of $i\chi$ at both sides, the heat flux can be obtained as
%$
%J={\langle}{\langle}I|\frac{{\partial}\mathcal{\hat{L}}_{\chi}}{{\partial}(i\chi)}|_{\chi=0}|\Psi{\rangle}{\rangle},
%$
%with $E_0(\chi=0)=0$, ${\langle}{\langle}I|\mathcal{\hat{L}}=0$ and $\mathcal{\hat{L}}=\hat{L}_{\chi}|_{\chi=0}$.
Here, we show full counting statistics of heat transfer at steady state by analytically exhibiting counting field based on the cumulant generating function ($G_{\chi}$). Since $G_{\chi}$ corresponds to the ground state energy ($E_0(\chi)=-G_{\chi}$), based on analysis in Appendix A, we obtain the ground eigen-solution in Liouville space, as
\begin{eqnarray}
E_0(\chi)&=&(X_e-X^{\chi}_e)
+\frac{Y-\sqrt{Y^2_{\chi}-(X^{\chi}_{o})^2+(X_o)^2}}{2}.\nonumber\\
\end{eqnarray}
The contributing term from the even  parity is
\begin{eqnarray}
X_e^{\chi}=\Gamma^{\chi}_{e,+}(0)+\Gamma^{\chi}_{e,-}(0),
\end{eqnarray}
and $X_e=X^{\chi}_e|_{\chi=0}$,
with the transition rate $\Gamma^{\chi}_{l,\sigma}(\omega)$ given at Eq.~(\ref{qme2}).
The terms from the odd parity are given by
\begin{eqnarray}
Y_{\chi}&=&\sum_{\sigma=\pm,\omega=\pm\Lambda}\Gamma^{\chi}_{o,\sigma}(\omega),\\
X^{\chi}_o&=&\sum_{\sigma=\pm,\omega=\pm\Lambda}{sgn(\omega)}{\sigma}\Gamma^{\chi}_{o,\sigma}(\omega), %\nonumber
\end{eqnarray}
with $sgn({\pm}\Lambda)={\pm}1$ and $\Lambda=\eta\Delta$.
$Y=Y_{\chi}|_{\chi=0}$ and
$X_{o}=X^{\chi}_o|_{\chi=0}$.
Consequently, the heat flux can be expressed as
\begin{eqnarray}~\label{current}
J&=&\frac{\Lambda^2}{8\pi}\int^{\infty}_{-\infty}
[\frac{\textrm{Re}[\Gamma_o(\Lambda)]C_o(-\Lambda,\omega')+\textrm{Re}[\Gamma_o(-\Lambda)]C_o(\Lambda,\omega')}
{\textrm{Re}[\Gamma_o(\Lambda)]+\textrm{Re}[\Gamma_o(-\Lambda)]}\nonumber\\
&&+C_e(0,\omega')]{\omega'}d{\omega'}
\end{eqnarray}
%{\color{blue}
where the rate probability densities are specified as
\begin{align}~\label{ceo1}
C_{e}(\omega,\omega^{\prime})&=\int^{\infty}_{-\infty}d\chi e^{-i\chi \omega'}\int^{\infty}_{-\infty}d\tau e^{i\omega\tau}[{\cosh}Q(\tau-\chi)-1],  \\ %\nonumber\\
C_{o}(\omega,\omega^{\prime})&=\int^{\infty}_{-\infty}d\chi e^{-i\chi \omega'}\int^{\infty}_{-\infty}d\tau e^{i\omega\tau}{\sinh}Q(\tau-\chi),
~\label{ceo2}
\end{align}
at energy $\omega=0, \pm\Lambda$. This analytical expression Eq.~(\ref{current}) of steady state heat flux without bias is found to be identical with the counterpart at Ref.~\cite{chenwang2015sp}, of which the turnover behavior on the coupling strength is  exhibited in Fig.~\ref{fig:fig3}(a) (blue dashed line). Physically, $C_{e}(0,\omega^{\prime})$ and $C_{o}(\pm\Lambda,\omega^{\prime})$ describe the even and odd parity components of the transfer process, respectively.
For example, $C_{o}(\Lambda,\omega^{\prime})$ describes the process that
the quit releases energy $\Lambda$ by relaxing from the excited eigen-state to the ground one, so that the right bath absorbs
energy $\omega'$ and the left one obtains the left $\Lambda-\omega'$. As such, the number of the state change of the qubit is odd, e.g., $n$ times excitation and $n+1$ times relaxation leads to a relaxation as the final action. And
$C_e(0, \omega')$ describes the process that the qubit has even number of virtual state-change, i.e., $n$ times relaxation and $n$ times excitation, so that the central qubit keeps intact and has no energy change. But still, the right bath absorbs
energy $\omega'$ and the left bath gains $-\omega'$ (i.e., releases $\omega'$).
%}
%{\color{red}
%which is identical with the expression of heat flux at Ref.~\cite{chenwang2015sp}.
%And the turnover behavior of heat flux is clearly shown in Fig.~\ref{fig:fig3}(a) (blue dashed line).
%%{\color{red}
%The rate probability densities shown at Eq.~(\ref{current}) are specified as
%\begin{align}~\label{ceo1}
%C_{e}(\omega,\omega^{\prime})&=\int^{\infty}_{-\infty}d\chi e^{-i\chi \omega'}\int^{\infty}_{-\infty}d\tau e^{i\omega\tau}[{\cosh}Q(\tau-\chi)-1],  \\ %\nonumber\\
%C_{o}(\omega,\omega^{\prime})&=\int^{\infty}_{-\infty}d\chi e^{-i\chi \omega'}\int^{\infty}_{-\infty}d\tau e^{i\omega\tau}{\sinh}Q(\tau-\chi),
%~\label{ceo2}
%\end{align}
%%with $C^{\pm}_{L}(\omega)=\int^{\infty}_{-\infty}d{\tau}e^{i\omega^{\prime}\tau{\pm}Q_v(\tau)}$ describing absorptions (emissions) of heat energy $\omega~(-\omega)$ in the $v$-th bath, which fulfills the detailed balance relation $C^{\pm}_v(\omega)/C^{\pm}_v(-\omega)=e^{\beta_v\omega}$, whereas $C_{e(o)}(\omega,\omega^{\prime})$ breaks this rule when the thermodynamic bias ($T_L{\neq}T_R$) of two baths is present.
%$C_{e(o)}(\omega,\omega^{\prime})$ shows the even (odd) parity contribution that as the two-level qubit relaxes from the excited state to the ground
%state by emitting energy $\omega$, the right phonon bath absorbs energy $\omega^{\prime}$ and the left bath gains $\omega-\omega^{\prime}$ if $\omega>\omega^{\prime}$ (or supplies the compensation $\omega^{\prime}-\omega$ if $\omega^{\prime}>\omega$).
%}

Similarly, the shot noise is obtained as
\begin{widetext}
\begin{align}~\label{shotnoise}
J^{(2)}&=\frac{\Lambda^2}{8\pi}\{
\int^{\infty}_{-\infty}{d\omega}[\frac{\textrm{Re}[\Gamma_o(-\Lambda)]C_o(\Lambda,\omega)+\textrm{Re}[\Gamma_o(\Lambda)]C_o(-\Lambda,\omega)}
{\textrm{Re}[\Gamma_o(\Lambda)]+\textrm{Re}[\Gamma_o(-\Lambda)]}%\nonumber\\
+C_e(0,\omega)]\omega^2 \nonumber \\%+\frac{(\phi_o(\Lambda)-\phi_o(-\Lambda))}{(\phi_o(\Lambda)+\phi_o(-\Lambda))^2}\nonumber\\
&-\frac{\int^{\infty}_{-\infty}{d\omega}(C_o(\Lambda,\omega)-C_o(-\Lambda,\omega))\omega}{(\textrm{Re}[\Gamma_o(\Lambda)]+\textrm{Re}[\Gamma_o(-\Lambda)])^3}%\nonumber\\
{\times}[\textrm{Re}[\Gamma_o(-\Lambda)]^2\int^{\infty}_{-\infty}\frac{d\omega}{\pi}C_o(\Lambda,\omega)\omega
-\textrm{Re}[\Gamma_o(\Lambda)]^2\int^{\infty}_{-\infty}\frac{d\omega}{\pi}C_o(-\Lambda,\omega)\omega]\}.%\nonumber\\
%&&{\times}[\int^{\infty}_{-\infty}\frac{d\omega}{\pi}(C_o(\Lambda,\omega)-C_o(-\Lambda,\omega))\omega]^2\}.
\end{align}
\end{widetext}
We find that the first term on the right side at Eq.~(\ref{shotnoise}) is the main contribution to the shot noise,
of which the spectral distribution is the same as that in heat flux at Eq.~(\ref{current}).
Hence, the nonmonotonic turnover behavior is quite similar to the heat flux, as shown in Fig.~\ref{fig:fig3}(b).

\subsubsection{Biased condition: $\epsilon_0\neq0$}
%{\color{blue}
Then, we extend analysis of steady state behaviors to the biased condition ($\epsilon_0\neq0$).
The heat flux shows the same nonmonotonic turnover behavior as $\alpha$ increases, i.e., the flux increases in the weak and moderate coupling strength regimes ($\alpha{\lesssim}1$) and decreases in the strong coupling regimes ($\alpha{\gtrsim}1$), shown in Fig.~\ref{fig:fig3}(a).
Interestingly, in the weak coupling regime ($\alpha{\lesssim}1$), the heat flux is enhanced by enlarging the qubit energy bias $\epsilon_0$, whereas as the coupling strength enters into the strong regime ($\alpha{\gtrsim}1$), the heat flux becomes intact for changing energy bias.
To confirm these results, we select typical coupling strengths to clearly demonstrate the influence of the energy bias on the heat flux, in Fig.~\ref{fig:fig3}(c).

Moreover, we look into the second-cumulant heat fluctuation, i.e., the noise power, at Fig.~\ref{fig:fig3}(b).
Similar to the steady state flux, the shot noise of heat flux also exhibits the same turnover behavior.
As the system-bath coupling strength increases, the noise power is enhanced by the energy bias in the weak coupling regime, whereas the noise power becomes nearly independent on the bias in the strong coupling regime.
Such behaviors are clearly depicted in Fig.~\ref{fig:fig3}(d).
Therefore, we conclude that both the steady state heat flux and the noise power are tuned in a similar way by either qubit-bath coupling or qubit energy bias.
%}
%{
%\color{red}
%Then, we extend calculations of steady state behaviors to the biased condition ($\epsilon_0\neq0$).
%By enlarging the qubit energy bias $\epsilon_0$, the heat flux shows nonmonotonic turnover behavior that increases in the weak and moderate coupling strength regimes ($\alpha{<}1$) and decreasing in the strong coupling regimes ($\alpha{>}1$), as shown in Fig.~\ref{fig:fig3}(a).
%As the interacting strength enters the strong coupling regime ($\alpha{\gtrsim}1$), the heat current becomes intact for changing energy bias.
%To confirm these results, we select typical coupling strengths to clearly demonstrate the influence of the energy bias on the heat flux, in Fig.~\ref{fig:fig3}(c).
%
%
%Moreover, we look into the higher-cumulant heat fluctuation, the noise power, at Fig.~\ref{fig:fig3}(b).
%Similar to the steady state flux, the shot noise of heat flux exhibits the same nonmonotonic turnover behavior as steady state heat flux. As the system-bath coupling strength increases, the qubit energy bias initially enhances the noise power, whereas the noise power becomes nearly independent on the energy bias in the strong coupling regime.
%Such behaviors are depicted in Fig.~\ref{fig:fig3}(d), which explicitly shows the relation of the steady state noise power with the energy bias.
%Therefore, we conclude that both the steady state heat flux and the noise power are tuned in a similar way by either qubit-bath coupling or qubit energy bias.
%}

\subsection{Geometric-phase induced heat flux}
As the system is periodically driven by external fields, e.g., modulated by two bath temperatures $T_{L(R)}(t)$, as schematically shown in Fig.~\ref{fig:fig0a},
the Liouville super-operator becomes time-dependent $\mathcal{\hat{L}}_{\chi}(t)$.
The effect of geometric phase will additionally contribute to the heat flux~\cite{nasinitsyn2007epl,nasinitsynprl2007,jieren2010prl,tianchen2013prb,tsagawa2011pre,tyuge2012prb}, demonstrated at Appendix B.
Thus, in the adiabatic modulation limit, there clearly exist two components to compose the generating function as
\begin{eqnarray}
\lim_{t{\rightarrow}\infty}\mathcal{Z}_{\chi}(t)=e^{{G}_{\chi}t}
=\exp{([{G}_{dyn}(\chi)+{G}_{geom}(\chi)]t)},
\end{eqnarray}
%{\color{red}
Specifically, the average dynamical phase is expressed as
${G}_{dyn}(\chi)=-\frac{1}{T_p}\int^{T_p}_0dtE_0(\chi,t)$,
where $T_p$ is the driving period, and $E_0(\chi,t)$ is the eigenvalue of $\mathcal{\hat{L}_{\chi}}(t)$ with the minimal real part.
It results in the dynamical heat flux
${J}_{dyn}=\frac{{\partial}}{{\partial}(i\chi)}{G}_{dyn}(\chi)|_{\chi=0}$.
The geometric phase contribution of the generating function is described by (Eq.~(\ref{gre}) at Appendix B)
\begin{eqnarray}~\label{geom1}~\label{ggeom}
{G}_{geom}(\chi)=-\frac{1}{T_p}\int^{T_p}_0dt{\langle}\Phi_{\chi}(t)|\frac{\partial}{{\partial}t}|\Psi_{\chi}(t){\rangle},
\end{eqnarray}
where $|\Psi_{\chi}(t){\rangle}~({\langle}\Phi_{\chi}(t)|)$ is the corresponding right (left) eigenvector of $E_0(\chi,t)$.
%}
Assuming two system parameters $u_1(t)$ and $u_2(t)$ are periodically modulated (which are two driving bath temperatures $T_{L(R)}(t)$ in this work), the geometric phase at Eq.~(\ref{geom1}) is specified as
${G}_{geom}(\chi)=-\frac{1}{T_p}\oint[du_1{\langle}\Phi_{\chi}|\frac{\partial}{{\partial}u_1}|\Psi_{\chi}{\rangle}
+du_2{\langle}\Phi_{\chi}|\frac{\partial}{{\partial}u_2}|\Psi_{\chi}{\rangle}]$.
According to the Stocks theorem, ${G}_{geom}(\chi)$ can be re-expressed as
\begin{eqnarray}~\label{geom2}
{G}_{geom}(\chi)=-\frac{1}{T_p}\int\int_{u_1,u_2}du_1du_2\mathcal{F}_{\chi}(u_1,u_2),
\label{eq:Gflux}
\end{eqnarray}
where %the Berry-like curvature is
\begin{eqnarray}
\mathcal{F}_{\chi}(u_1,u_2)={\langle}{\partial}_{u_{1}}\Phi_{\chi}|{\partial}_{u_{2}}\Psi_{\chi}{\rangle}
-{\langle}{\partial}_{u_{2}}\Phi_{\chi}|{\partial}_{u_{1}}\Psi_{\chi}{\rangle}.
\end{eqnarray}

%{\color{red}
It is worth noting~\cite{jieren2012prl} that $\mathcal{F}_{\chi}(u_1,u_2)$ has the meaning of curvature
in the parameter space $(u_1, u_2)$ of the ground state of quantum Liouville super-operator $\mathcal{\hat{L}_{\chi}}$. It is of pure geometric interpretation and independent of the driving
speed (in adiabatic limit). Mathematically, ${G}_{geom}(\chi)$ is an analog of the adiabatic
Berry phase in quantum mechanics~\cite{mberry1984prla}, where in the latter case the wave
function will obtain an extra phase after a cyclic evolution. Similarly, in the full counting statistics of our
driven systems, the cumulant generating function ${G}_{geom}(\chi)$  (analog
of phase) in the exponent of the characteristic function $\mathcal{Z}_{\chi}$
(analog of wave function) will obtain also an additional
term. Both extra terms share the similar geometric
origin from the nontrivial curvature in the system's parameter
space. As such $\mathcal{F}_{\chi}(u_1,u_2)$ is a Berry-like curvature and we term ${G}_{geom}(\chi)$ the geometric phase contribution, which can generates the $n$-th cumulant of geometric phase induced heat current fluctuation, as~\cite{jieren2010prl, tianchen2013prb, nasinitsyn2007epl}
%}
\begin{eqnarray}~\label{geom3}
{J}^{(n)}_{geom}&=&\frac{{\partial}^n{G}_{geom}(\chi)}{{\partial}(i\chi)^n}|_{\chi=0}\\
&=&-\frac{1}{T_p}\int\int_{u_1,u_2}du_1du_2\frac{\partial^n}{{\partial}(i\chi)^n}\mathcal{F}_{\chi}(u_1,u_2)|_{\chi=0}. \nonumber
\end{eqnarray}
The geometric heat flux is given by the first cumulant ${J}_{geom}={J}^{(1)}_{geom}$.

\subsubsection{Unbiased condition: $\epsilon_0=0$}
Here, we firstly investigate the geometric heat flux without bias ($\epsilon_0=0$).
As already known that in the weak qubit-bath coupling regime, the geometric-phase induced heat flux is finite and
independent on the coupling strength~\cite{jieren2010prl}. This mainly results from the fact that with weak qubit-bath coupling the transition rates between the two-level qubit and phononic baths are linearly dependent on the coupling strength, exhibiting the additive transfer processes.
On the contrary, the geometric heat flux vanishes in the strong qubit-bath coupling regime by applying the nonequilibrium NIBA method~\cite{tianchen2013prb}. The left and right eigenvectors corresponding to the ground state energy
are given by $|\Psi_{\chi}{\rangle}=\frac{1}{2}[1,1,0,0]^{T}$ and
${\langle}\Phi_{\chi}|=[1,1,0,0]$, which are clearly independent of the system parameters and result in the zero geometric heat flux according to Eq.~(\ref{eq:Gflux}).
It was proposed that these two approaches describe different physical pictures within the same NESB system,
and was not conflicted with each other~\cite{tianchen2013prb,chenwang2015sp}.

\begin{figure}[tbp]
\begin{center}
%\vspace{-0.5cm}
\includegraphics[scale=0.42]{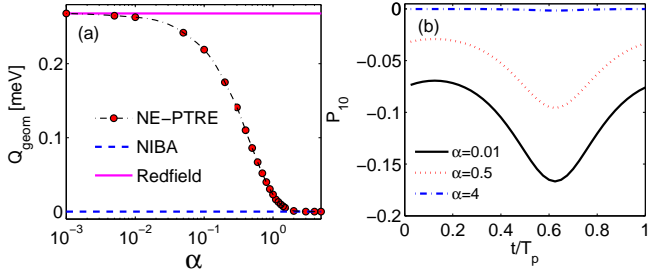}
\vspace{-1.0cm}
\end{center}
\caption{(Color online) Adiabatic modulation by two bath temperatures without bias ($\epsilon_0=0$):
(a) geometric phase induced heat pump $Q_{geom}={J}_{geom}{\times}T_p$;
(b) the coherence ($P_{10}$) in local basis.
Two bath temperatures are specified as $T_L(\tau)=(150+90\cos\Omega_p\tau)$~K and
$T_R(\tau)=(150+90\sin\Omega_p\tau)$~K, with the period $T_p=1$~ns.
The other parameters are given by $\Delta=5.22$~meV, $\omega_c=26.1$~meV.
}~\label{fig:fig1}
\end{figure}

Based on the $\chi$-dependent NE-PTRE at Eq.~(\ref{qme2}), we try to explicitly unify these limiting results,
as shown in Fig.~\ref{fig:fig1}(a).
In the weak system-bath coupling regime, the geometric heat flux approaches to the upper limit within the Redfield scheme.
As the coupling strength increases, geometric heat flux is strongly suppressed and asymptotically decreases to zero,
which finally becomes identical with the result in the nonequilibrium NIBA.
The underlying mechanism can be understood by analyzing the coherence $P_{10}(t)$, since the populations ($P_{00}, P_{11}$) are constant.
We find in Fig.~\ref{fig:fig1}(b)  that the coherence is suppressed monotonically
by increasing the qubit-bath coupling strength, finally resulting in the constant quasi-steady state in the strong coupling limit~\cite{tianchen2013prb}.
It is proposed that without bias ($\epsilon_0=0$), multi-phonons involved processes deteriorate
the construction of geometric phase.
Therefore, these seemingly contradiction is clearly solved within the framework of NE-PTRE accompanied with the counting field.

%%==========================================
%%==========================================

Moreover, compared to the dynamical heat flux~\cite{chenwang2015sp}, the system-bath coupling plays
a distinct role in the geometric heat flux.
For the dynamical flux,
in the weak and intermediate coupling regimes, multi-phonon involved processes are helpful to generate steady state heat flux,
mainly due to the robustness of transition rates. While in the strong coupling limit, the large system-bath interaction weakens the transition rates due to quantum Zeno-like effect, and finally suppresses the heat flux.
Hence, the non-monontonic behavior of the dynamical heat flux is clearly demonstrated.
For the geometric flux, increasing the system-bath coupling strength will only monotonically decrease the geometric heat flux,
which implies that the instantaneous state of the qubit are inclined to stay intact, which is independent of temperature modulations, as we discussed above.
%{\color{red}(shown as the dashed-dot blue line in Fig.~\ref{fig:resonance}(a))}.

%%==========================================
%\begin{figure}[tbp]
%\begin{center}
%\vspace{-2.2cm}
%\includegraphics[scale=0.43]{fig3a.eps}
%\includegraphics[scale=0.45]{fig3b.eps}
%\includegraphics[scale=0.45]{fig3c.eps}
%\vspace{-2.2cm}
%\end{center}
%\caption{(Color online)
%Adiabatic modulation by two bath temperatures under finite energy bias  of the qubit ($\epsilon_0=2.61$~meV):
%(a) geometric phase induced heat pump $J_{geom}*T_p$;
%(b) Population ($P_{11}$) and real part of coherence ($\textrm{Re}(P_{10}$);
%(c) effective rates of the population ($T_p*\frac{dP_{11}}{dt}$)
%and the coherence ($T_p*\frac{d\textrm{Re}(P_{10})}{dt}$).
%While other parameters are the same as in Fig.~\ref{fig:fig1}.
%}~\label{fig:bias}
%\end{figure}
%%==========================================

\subsubsection{Biased condition: $\epsilon_0\neq0$}
\begin{figure}[tbp]
\begin{center}
%\vspace{-2.2cm}
\includegraphics[scale=0.4]{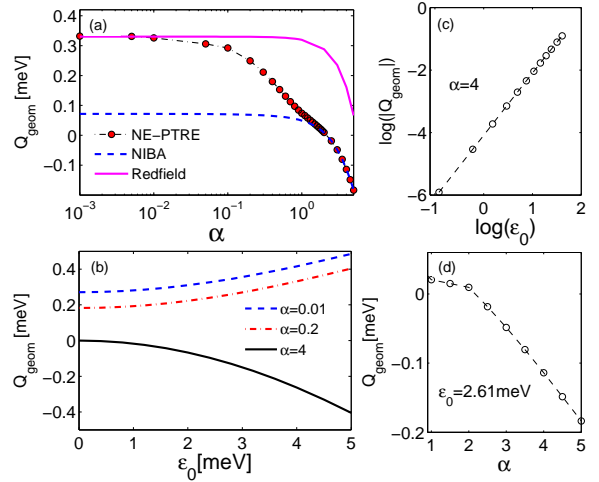}
%\vspace{-2.2cm}
\end{center}
\caption{(Color online)
(a) Geometric phase induced heat pump $Q_{geom}=J_{geom}{\times}T_p$ under finite energy bias  of the qubit ($\epsilon_0=2.61$~meV);
(b) influence of qubit energy bias on the geometric-phase induced heat pump, by modulating two bath temperatures;
(c) the log-log relation between $\epsilon_0$ and $Q_{geom}$ in strong coupling regime ($\alpha=4$);
(d) the linear relation between $Q_{geom}$ and $\alpha$ in strong coupling regime.
The parameters are the same as in Fig.~\ref{fig:fig1}.
}~\label{fig:fig2}
\end{figure}

Next, we analyze the geometric heat flux under finite energy bias ($\epsilon_0{\neq}0$), as shown in Fig.~\ref{fig:fig2}(a).
In the weak coupling limit, the geometric heat flux is equal to that from the Redfield scheme.
The existence of coherence $P_{10}$  is also crucial to enhance the geometric-phase induced heat flux, which is similar to the unbiased case in
Fig.~\ref{fig:fig1}.
As the coupling strength increases, the geometric heat flux decreases sharply, and even turns to be negative.
The corresponding coherence is strongly suppressed,
which leaves only the populations to contribute to the geometric heat flux.
Then, the behavior of geometric heat flux is consistent with the result within nonequilibrium NIBA in the strong coupling regime~\cite{tianchen2013prb}.
As a result, we conclude that NE-PTRE can also be applicable to unify limiting coupling results beyond unbiased condition.

%%==========================================
%%==========================================

%% effect of zeeman energy on the geometric heat pump
Then, we turn to analyze the influence of qubit energy bias on geometric heat pump at Fig.~\ref{fig:fig2}(b).
In the weak qubit-bath coupling regime (e.g., $\alpha=0.01$), geometric heat pump shows monotonic enhancement by increasing the
energy bias.
As the interaction strength is modulated to the intermediate coupling regime (e.g., $\alpha=0.2$),
geometric heat pump is also positively enhanced by the increasing energy bias, which is similar to the counterpart in the weak coupling case.
If we further enlarge the coupling strength (e.g., $\alpha=4$), the geometric heat pump becomes negatively enhanced, which is quantitatively distinct from that in the weak coupling regime.
This observation clearly demonstrates different physical pictures in these two limiting interaction regimes.

%{\color{red}(analytical solution for biased case in strong coupling limit)}

We should admit that it is out of our ability to analytically give a comprehensive picture in a wide system-bath coupling regime for the biased case.
Here, to understand the geometric heat flux reversal, we focus on the strong interaction limit that is consistent with the nonequilibrium NIBA framework.
Combined with the counting filed, the equation of motion for the qubit is expressed as
\begin{equation}
\frac{d}{dt}
\begin{pmatrix}
P^{\chi}_{11}\\
P^{\chi}_{00}\\
\end{pmatrix}
=-
\begin{pmatrix}
K(\epsilon_0) & -K_{-}(\chi)\\
-K_+(\chi) & K(-\epsilon_0)\\
\end{pmatrix}
\begin{pmatrix}
P^{\chi}_{11}\\
P^{\chi}_{00}\\
\end{pmatrix}
\end{equation}
with the population $P^{\chi}_{ii}={\langle}i|\hat{\rho}_{\chi}(t)|i{\rangle}$.
%{\color{red}
The transition rates are given by
\begin{eqnarray}
K^{\pm}(\chi)=(\Delta/2)^2\int^{\infty}_{-\infty}dt\eta^2e^{{\pm}i\epsilon_0t+Q_L(t)+Q_R(t-\chi)},
\end{eqnarray}
with $K({\pm}\epsilon_0)=K_{\pm}(\chi)|_{\chi=0}$, $\eta$ and $Q_v(t)$ given at Eq.~(\ref{eta}) and Eq.~(\ref{qvf}), respectively.
%}
Thus, the eigen-state energies are directly obtained as
\begin{align}
E_{\pm}(\chi)&=\frac{1}{2}[(K(\epsilon_0)+K(-\epsilon_0)){\pm}\nonumber\\
&\sqrt{(K(\epsilon_0)-K(-\epsilon_0))^2+4K_{+}(\chi)K_{-}(\chi)}].
\end{align}
The corresponding right eigen-states are given by
\begin{eqnarray}
|\Psi^{\pm}_\chi{\rangle}=[2K_{-}(\chi),A_{\pm}(\chi)]^{T}
\end{eqnarray}
with the coefficients
$A_{\pm}(\chi)=(K(\epsilon_0)-K(-\epsilon_0)){\mp}\sqrt{(K(\epsilon_0)-K(-\epsilon_0))^2+4K_{+}(\chi)K_{-}(\chi)}$.
Accordingly, the  left eigen-states are
\begin{eqnarray}
\langle\Phi^{\pm}_\chi|=\frac{[2K_{+}(\chi),A_{\pm}(\chi)]}{A^2_{\pm}(\chi)+4K_+(\chi)K_-(\chi)}.
\end{eqnarray}

%{\color{red}
In the strong qubit-bath coupling limit, it is known that the Marcus approximation becomes applicable~\cite{dvirasegal2006prb,rmarcus1956jcp}.
Marcus theory was originally proposed to the electron transfer rate in the donor-acceptor species.
And it works at high temperature $k_BT{>}\epsilon_0$ and/or the strong qubit-bath coupling regime~\cite{refereeQ}.
It can be approached by the short-time expansion of $Q_v(t)$ at Eq.~(\ref{qvf}) as
$Q_v(t)=\frac{\Gamma_vT_v}{\omega^2_{c,v}}-\Gamma_vT_vt^2-i\Gamma_vt$~\cite{note2}, with the effective coupling strength
$\Gamma_v=\int\frac{J_v(\omega)}{\pi\omega}d\omega=2\alpha_v\omega_{c,v}$.
%}
Consequently, the transition rates combined with the counting parameter are simplified as
$K_{\pm}(\chi)=K(\pm\epsilon_0)M_{\pm}(\chi)$,
with the standard rates
\begin{eqnarray}~\label{kpm}
K(\pm\epsilon_0)&=&\frac{\Delta^2}{4}\sqrt{\frac{\pi}{\Gamma_LT_L+\Gamma_RT_R}}
\exp[-\frac{(\epsilon_0{\mp}\Gamma_L{\mp}\Gamma_R)^2}{4(\Gamma_LT_L+\Gamma_RT_R)}].\nonumber\\
\end{eqnarray}
and the factor
\begin{equation}
M_{\pm}(\chi)=e^{{\pm}i\epsilon_0\chi-\frac{\Gamma_LT_L\Gamma_RT_R}{\Gamma_LT_L+\Gamma_RT_R}
[i\chi(\frac{1}{T_L}+\frac{\pm\epsilon_0-\Gamma_R}{\Gamma_RT_R})+\chi^2]}.
\end{equation}
%{\color{red}
In absence of the counting field ($\chi=0$), the factor $M_{\pm}(\chi=0)=1$,
and the modified transition rates $K_{\pm}(\chi)$ reduce back to the standard expressions $K(\pm\epsilon_0)$ at Eq.~(\ref{kpm}), respectively.
%}
Moreover, we consider the weak qubit energy bias regime, i.e., $\epsilon_0{\ll}\{\Gamma_v,k_BT_v\}$. Then the transition rate
at Eq.~(\ref{kpm}) can be approximately expanded up to the first order of $\epsilon_0$ as
\begin{eqnarray}
K({\pm}\epsilon_0){\approx}K_0[1{\pm}\frac{\epsilon_0}{2(\Gamma_LT_L+\Gamma_RT_R)}(\Gamma_L+\Gamma_R)],
\end{eqnarray}
with
$K_0=\frac{\Delta^2}{4}\sqrt{\frac{\pi}{\Gamma_LT_L+\Gamma_RT_R}}
\exp[-\frac{(\Gamma_L+\Gamma_R)^2}{4(\Gamma_LT_L+\Gamma_RT_R)}]$.
According to the definition at Eq.~(\ref{geom3}),
the geometric phase induced heat flux is obtained as
\begin{equation}~\label{jgeoma}
J_{geom}=-\frac{\epsilon^2_0}{T_p}\int\int_{T_L,T_R}dT_LdT_R\frac{\Gamma_L\Gamma_R(\Gamma_L+\Gamma_R)^3}
{8(\Gamma_LT_L+\Gamma_RT_R)^4}.
\end{equation}
This expression clearly confirms the reversal (negative) behavior of the heat flux in strong coupling limit, shown in Fig.~\ref{fig:fig2}(a).
Moreover, the power-law feature of energy bias is analytically exhibited at Eq.~(\ref{jgeoma}),
which is excellently agreeable with the numerical result as $J_{geom}{\sim}-\epsilon^{2.0{\pm}0.02}_0$, shown in Fig.~\ref{fig:fig2}(c).
If the system-bath couplings are identically selected as $\alpha_L=\alpha_R=\alpha$,
the geometric heat flux is expressed as $J_{geom}{\sim}-\alpha\epsilon^2_0$  based on Eq.~(\ref{jgeoma}),
and numerically confirmed at Fig.~\ref{fig:fig2}(d),
which coincides with the numerical results of Fig.~2 in Ref.~\cite{tianchen2013prb} that $J_{geom}$ is linear-dependent on the coupling strength $\alpha$ and quadratic-dependent on the qubit energy bias $\epsilon_0$.

\section{Conclusion}
In summary, we have investigated the geometric-phase induced heat pump in the nonequilibrium spin-boson model by periodically modulating temperatures of two thermal baths, which is beyond the traditional Redfield and nonequilibrium NIBA schemes.
By developing nonequilibrium polaron-transformed Redfield equation (NE-PTRE) approach in the context of full counting statistics, the cumulant generating function is clearly demonstrated,
which consists of both the dynamical phase and geometric phase contributions.
In absence of the external driving field, the influences of qubit energy bias on the steady state heat flux and the corresponding noise power have been analyzed.
When in the weak and moderate coupling regimes, the energy bias monotonically enhances both the steady state heat flux and the
noise power. While in the strong coupling regime, these two observables become independent on the energy bias.
This clearly demonstrates the same role of the energy bias in affecting the heat flux and the noise power.

Then, we have analyzed geometric heat pump without bias by varying qubit-bath coupling strength in a wide regime.
In the weak system-bath coupling limit, geometric heat flux is positive finite, which is equivalent to the counterpart within Redfield scheme~\cite{jieren2010prl}. As the coupling strength increases, geometric heat flux shows monotonic decrease, and
finally approaches strictly zero, which is identical with the result based on nonequilibrium NIBA~\cite{tianchen2013prb}.
We have also studied the geometric heat pump at the biased condition.
We found that geometric heat pump decreases quickly as the qubit-bath coupling increases,
and shows reversal behavior in the strong coupling regime.
Moreover, the analytical relations of the geometric heat flux with the system-bath coupling and the energy bias have been obtained.
The results based on the NE-PTRE also show consistency with the counterparts from the Redfield and nonequilibrium NIBA schemes,
in the weak and strong coupling regimes, respectively.

Therefore, we conclude that this unified theory is applicable to obtain the geometric heat flux in the nonequilibrium spin-boson model,
both at unbiased and biased conditions.
Moreover, we have turned to analyze the influence of the qubit energy bias on the geometric heat pump.
Geometric heat flux is negatively enhanced in the strong qubit-bath coupling regime, which is in sharp contrast to
the counterpart in the weak coupling case, exhibiting positive stabilization.
We hope that these results would have broad implications for smart control of energy transfer in low-dimensional nanodevices.

\section{acknowledgement}
Chen Wang is supported by the National Natural Science Foundation of China under Grant Nos. 11547124 and 11574052.
Jie Ren acknowledges the National Youth 1000 Talents Program in China, and the 985 startup grant (No. 205020516074) at Tongji University.

\appendix

\section{Analytical expression of the steady state cumulant generating function without bias}

Without bias ($\epsilon_0=0$),
the Liouvillian dynamics of the reduced density matrix at Eq.~(\ref{lde}) under the framework of full counting statistics is expressed as
$\frac{d}{dt}|\rho_{\chi}{\rangle}=-\hat{L}_{\chi}{|\rho_{\chi}{\rangle}}$, where the evolution matrix is specified as
\begin{eqnarray}
\hat{L}_{\chi}=
\begin{pmatrix}
a & -a_{\chi} & b_{\chi} & c_{\chi} \\
-a_{\chi} & a & c_{\chi} & b_{\chi} \\
d_{\chi} & e_{\chi} & a & f_{\chi} \\
e_{\chi} & d_{\chi} & f_{\chi} & a\\
\end{pmatrix}.
\end{eqnarray}
The matrix elements are shown as
$a_{\chi}=X^{\chi}_e+\frac{Y_{\chi}}{2}$,
$b_{\chi}=-\frac{1}{2}(X^{\chi}_{o,+}+X_{o,-})$,
$c_{\chi}=\frac{1}{2}(X_{o,+}+X^{\chi}_{o,-})$,
$d_{\chi}=\frac{1}{2}(X^{\chi}_{o,+}-X_{o,-})$,
$e_{\chi}=\frac{1}{2}(X_{o,+}-X^{\chi}_{o,-})$,
$f_{\chi}=-X^{\chi}_e+\frac{Y_{\chi}}{2}$,
and $a=a_{\chi}|_{\chi=0}$,
with the coefficients
\begin{eqnarray}
X^{\chi}_e=\Gamma^{\chi}_{e,+}(0)+\Gamma^{\chi}_{e,-}(0),
\end{eqnarray}
\begin{eqnarray}
Y_{\chi}=\Gamma^{\chi}_{o,+}(\Lambda)+\Gamma^{\chi}_{o,+}(-\Lambda)+\Gamma^{\chi}_{o,-}(\Lambda)+\Gamma^{\chi}_{o,-}(-\Lambda),
\end{eqnarray}
\begin{eqnarray}
X^{\chi}_{o,\pm}=\Gamma^{\chi}_{o,\pm}(\Lambda)-\Gamma^{\chi}_{o,\pm}(-\Lambda),
\end{eqnarray}
and $X_{o,\pm}=X^{\chi}_{o,\pm}|_{\chi=0}$.
The modified transition rates $\Gamma^{\chi}_{e(o)}(\omega)$ are shown at Eq.~(\ref{qme2}).

To find the eigen-values of the evolution matrix, we set
$\textrm{det}(\textrm{L}_{\chi}-\lambda\textrm{I})=0$, which results in
\begin{eqnarray}
(a-\lambda)^2&=&(a_{\chi}f_{\chi}+b_{\chi}d_{\chi}+c_{\chi}e_{\chi})\\
&&{\pm}[(a_{\chi}-f_{\chi})(a-\lambda)+(c_{\chi}d_{\chi}+b_{\chi}e_{\chi})].\nonumber
\end{eqnarray}
For one branch, the solution is given by
\begin{eqnarray}
\lambda^p_{\pm}&=&(X_e-X^{\chi}_e)+\frac{Y}{2}\\
&&{\mp}\sqrt{Y^2_{\chi}-(X^{\chi}_{o,+}-X^{\chi}_{o,-})^2+(X_{o,+}-X_{o,-})^2}/2,\nonumber
\end{eqnarray}
And for the other branch, it is given by
\begin{eqnarray}
\lambda^m_{\pm}&=&(X_e+X^{\chi}_e)+\frac{Y}{2}\\
&&{\mp}\sqrt{Y^2_{\chi}-(X^{\chi}_{o,+}+X^{\chi}_{o,-})^2+(X_{o,+}+X_{o,-})^2}/2\nonumber
\end{eqnarray}

Hence, the ground state energy is obtained as $E_0(\chi)=\lambda^p_+$.
Since the cumulant generating function is given by $G_{\chi}=-E_0(\chi)$, it is specified as
\begin{eqnarray}
{G}_{\chi}=(X^{\chi}_e-X_e)-\frac{Y}{2}
{+}\sqrt{Y^2_{\chi}-(X^{\chi}_o)^2+(X_o)^2}/2,
\end{eqnarray}
with $X^{\chi}_o=X^{\chi}_{o,+}-X^{\chi}_{o,-}$ and $X_o=X^{\chi}_o|_{\chi=0}$.

\section{Introduction of geometric phase and cumulant generating function}

Considering the time-dependent super-operator $\hat{L}_{\chi}(t)$ with the counting parameter, which is not hermitian,
we obtain the quasi-eigen solution as
\begin{eqnarray}
\hat{L}_{\chi}(t)|\psi_n(\chi,t){\rangle}&=&E_n(\chi,t)|\psi_n(\chi,t){\rangle},\\
{\langle}\phi_n(\chi,t)|\hat{L}_{\chi}(t)&=&{\langle}\phi_n(\chi,t)|E_n(\chi,t),\nonumber
\end{eqnarray}
where $\lambda_n(\chi,t)$ is the instantaneous eigenvalue of $\hat{L}_{\chi}(t)$,
and $|\psi_n(\chi,t){\rangle}$ (${\langle}\phi_n(\chi,t)|$) is the corresponding
normalized right (left) eigenvector, which obeys the relation ${\langle}\phi_n(\chi,t)|\psi_n(\chi,t){\rangle}=\delta_{n,m}$.
In analogy with the seminal Berry's solution, we can express the wavefunction in the basis
$\{|\psi_n(\chi,t)\}$ as
\begin{eqnarray}~\label{awavefunction}
|\rho_{\chi}(t){\rangle}=\sum_na_n(t)\exp[-\int^{t}_0E_n(\chi,\tau)d{\tau}]|\psi_n(\chi,t){\rangle},
\end{eqnarray}
By substituting Eq.~(\ref{awavefunction}) into the dynamical equation at Eq.~(\ref{lde}),
we obtain the evolution equation of $a_n(t)$
\begin{eqnarray}~\label{aan}
&&\sum_n\frac{d{a}_n(t)}{dt}\exp[-\int^{t}_0E_n(\chi,\tau)d\tau]|\psi_n(\chi,t){\rangle}\nonumber\\
=&&-\sum_na_n(t)
\exp[-\int^{t}_0E_n(\chi,\tau)d\tau]|\frac{d}{dt}{\psi}_n(\chi,t){\rangle}.
\end{eqnarray}

Then, by left-multiplying the eigenvector ${\langle}\phi_m(\chi,t)|$ to Eq.~(\ref{aan}), we find that
\begin{eqnarray}~\label{adan}
\frac{d{a}_m(t)}{dt}&=&-a_m(t){\langle}\phi_m(\chi,t)|\frac{d}{dt}{\psi}_{m}(\chi,t){\rangle}\nonumber\\
&&-\sum_{n{\neq}m}a_n(t)\exp(-\int^t_0[E_n(\chi,\tau)-E_m(\chi,\tau)]d\tau)\nonumber\\
&&{\times}{\langle}\phi_m(\chi,t)|\frac{d}{dt}{\psi}_{n}(\chi,t){\rangle}.
\end{eqnarray}
It should be noted that the eigenvalue $E_n(\chi,t)$ generally is the complex value.
Hence, the long time behavior of the reduced qubit system is mastered by only the eigenmode $m=0$,
of which the eigenvalue $E_0(\chi,t)$ owns the smallest real part.

In the adiabatic limit, the second term at the right side of the Eq.~(\ref{adan}) can be approximately ignored due to the decay factor $\exp(-\int^t_0[E_n(\chi,\tau)-E_0(\chi,\tau)]d\tau)$~($\textrm{Re}[E_n(\chi,\tau)-E_0(\chi,\tau)]>0$ for $n{\neq}0$).
We obtain the expression of $a_n(t)$ after long time evolution ($t{\rightarrow}\infty$) as
\begin{eqnarray}
a_0(t)=\exp(-\int^{t}_0{\langle}\phi_0(\chi,\tau)|\frac{d}{d\tau}\psi_0(\chi,\tau){\rangle}d\tau)a_0(0),
\end{eqnarray}
with $a_0(0)$ the initial state coefficient.
Then, if we consider the adiabatic cyclic evolution over a long time period $T_p$,
the wave function can be specified as
\begin{eqnarray}
|\rho_{\chi}(t){\rangle}&=&\exp(-\frac{t}{T_p}\int^{T_p}_0
d\tau[E_0(\chi,\tau)\\
&&+{\langle}\phi_0(\chi,\tau)|\frac{d}{d\tau}\psi_0(\chi,\tau){\rangle}])a_0(0)
|\rho_{\chi}(0){\rangle}.\nonumber
\end{eqnarray}
Consequently, the generating function can be obtained as
\begin{eqnarray}
\mathcal{Z}_{\chi}(t)&=&{\langle}\textrm{I}|\rho_{\chi}(t){\rangle}\\
&{\approx}&\exp(-\frac{t}{T_p}\int^{T_p}_0d\tau
[E_0(\chi,\tau)\nonumber\\
&&+{\langle}\phi_0(\chi,\tau)|\frac{d}{d\tau}\psi_0(\chi,\tau){\rangle}])a_0(0)\nonumber
{\langle}\textrm{I}|\rho_{\chi}(0){\rangle}.\nonumber
\end{eqnarray}
Finally, the cumulant generating function in long time limit can be described by two contributing terms as
\begin{eqnarray}
G(\chi)=\lim_{t{\rightarrow}\infty}\frac{\ln\mathcal{Z}_{\chi}(t)}{t}=G_{dyn}(\chi)+G_{geom}(\chi),
\end{eqnarray}
and the factor $\lim_{t{\rightarrow}\infty}\frac{1}{t}\ln(a_0(0){\langle}\textrm{I}|\rho_{\chi}(0){\rangle})$
becomes negligible.
Here, $G_{dyn}(\chi)$ is the dynamical phase factor, shown as
$G_{dyn}(\chi)=-\frac{1}{T_p}\int^{T_p}_0E_0(\chi,\tau)d\tau$.
While $G_{dyn}(\chi)$ originates from the geometric phase contribution, shown as
$G_{geom}(\chi)=-\frac{1}{T_p}\int^{T_p}_0{\langle}\phi_0(\chi,\tau)|\frac{d}{d\tau}\psi_0(\chi,\tau){\rangle}d\tau$.
In the main context, we use $|\Psi_{\chi}(t){\rangle}~({\langle}\Phi_{\chi}(t)|)$ at Eq.~(\ref{ggeom})
to replace $|\psi_0({\chi},t){\rangle}~({\langle}\phi_{0}(\chi,t)|)$.
The geometric phase induced cumulant generating function is re-expressed as
\begin{eqnarray}~\label{gre}
G_{geom}(\chi)=-\frac{1}{T_p}\int^{T_p}_0{\langle}\Phi_{\chi}(\tau)|\frac{d}{d\tau}\Psi_{\chi}(\tau){\rangle}d\tau
\end{eqnarray}

%\section{appendix: Nonequilibrium polaron transformed Redfield equation combined with counting field parameter}


\begin{thebibliography}{99}
\bibitem{ydubi2011rmp} Y. Dubi and M. Di Ventra, Rev. Mod. Phys. \textbf{83}, 131 (2011).
\bibitem{maratner2013nn} M. A. Ratner, Nat. Nanotechnology \textbf{8}, 378 (2013).
\bibitem{pnalbach2013pnas} P. Nalbach and M. Thorwart, PNAS \textbf{8}, 2693 (2013).
\bibitem{MMohseni2014book} M. Mohseni, Y. Omar, G. S. Engel, and M. B. Plenio,
\emph{Quantum Effects in Biology}. (Cambridge University Press, United Kingdom, 2014).
\bibitem{dzxu2016fp} D. Z. Xu and J. S. Cao, Front. Phys. \textbf{11}, 110308 (2016).
\bibitem{nbli2012rmp} N. B. Li, J. Ren, L. Wang, G. Zhang, P. H\"{a}nggi, and B. W. Li,
Rev. Mod. Phys. \textbf{84}, 1045 (2012).


\bibitem{jren2015aip} J. Ren and B. W. Li, AIP Advances \textbf{5}, 053101 (2015).
\bibitem{dvirasegal2005prl} D. Segal and A. Nitzan, Phys. Rev. Lett. \textbf{94}, 034301 (2005).
\bibitem{dvirasegal2006prb} D. Segal, Phys. Rev. B \textbf{73}, 205415 (2006).
\bibitem{cwwang2006science} C. W. Chang, D. Okawa, A. Majumdar, and A. Zettl, Science \textbf{314}, 1121 (2006).
\bibitem{lwangprl2007} L. Wang and B. W. Li, Phys. Rev. Lett. \textbf{99}, 177208 (2007).
\bibitem{lfzhang2010prl} L. F. Zhang, J. Ren, J. S. Wang, and B. W. Li, Phys. Rev. Lett. \textbf{105}, 225901 (2010).
\bibitem{etaylor2015prl} E. Taylor and D. Segal, Phys. Rev. Lett. \textbf{114}, 220401 (2015).

\bibitem{dsegal2008prl} D. Segal, Phys. Rev. Lett. \textbf{101}, 260601 (2008).
\bibitem{jieren2010prl} J. Ren, P. Hanggi, and B. W. Li, Phys. Rev. Lett. \textbf{104}, 170601 (2010).
\bibitem{tianchen2013prb} T. Chen, X. B. Wang, and Jie Ren, Phys. Rev. B \textbf{87}, 144303 (2013).
\bibitem{cuchiyama2014pre} C. Uchiyama, Phys. Rev. E \textbf{89}, 052108 (2014).
%\bibitem{jieren2012prl} J. Ren, S. Liu, and B. W. Li, Phys. Rev. Lett. \textbf{108}, 210603 (2012).

\bibitem{ksaito2013prl} K. Saito and T. Kato, Phys. Rev. Lett. \textbf{111}, 214301 (2013).
\bibitem{ajleggett1987rmp} A. J. Leggett, S. Chakravarty, A. T. Dorsey, M. P. A. Fisher, A. Garg, W. Zwerger,
Rev. Mod. Phys. \textbf{59}, 1 (1987).
\bibitem{uweiss2008book} U. Weiss, \emph{Quantum Dissipative Systems} (World Scientific, Singapore, 2008).


\bibitem{lnicolinjcp2011} L. Nicolin and D. Segal, J. Chem. Phys. \textbf{135}, 164106 (2011).
\bibitem{lnicolinprb2011} L. Nicolin and D. Segal, Phys. Rev. B \textbf{84}, 161414 (2011).
\bibitem{dsegalpre2014} D. Segal, Phys. Rev. E \textbf{90}, 012148 (2014).


\bibitem{kavelizhanin2008cpl} L. A. Velizhanin, H. Wang and M. Thoss, Chem. Phys. Lett. \textbf{460}, 325 (2008).

\bibitem{junjieliu2016a} J. J. Liu, H. Xu and C. Q. Wu, arXiv.1608.04854.
\bibitem{junjieliu2016b} J. J. Liu, H. Xu, B. W. Li and C. Q. Wu, arXiv.1609.05598.

\bibitem{bkagarwalla1612} B. K. Agarwalla and D. Segal, arXiv:1612.01008.

\bibitem{chenwang2015sp} C. Wang, J. Ren and J. S. Cao, Sci. Rep. \textbf{5}, 11787 (2015).

\bibitem{sgasparinetti2014njp} S. Gasparinetti, P. Solinas, A. Braggio and M. Sassetti, New J. Phys. \textbf{16}, 115001 (2014).
\bibitem{gguarnieri2016pra} G. Guarnieri, C. Uchiyama and B. Vacchini, Phys. Rev. A \textbf{93}, 012118 (2016).
\bibitem{jcerrillo2016arxiv} J. Cerrillo, M. Buser and T. Brandes, arXiv.1606.05074.
\bibitem{mcarrega2016prl} M. Carrega, P. Solinas, M. Sassetti and U. Weiss, Phys. Rev. Lett. \textbf{116}, 240403 (2016).
\bibitem{lferialdi1609} L. Ferialdi, arXiv:1609.00645.

\bibitem{djthouless1983prb} D. J. Thouless, Phys. Rev. B \textbf{27}, 6083 (1983).
\bibitem{anazir2009prl} A. Nazir, Phys. Rev. Lett. \textbf{103}, 146404 (2009).

\bibitem{sjang2008jcp} S. Jang, Y. C. Cheng, D. R. Reichman and J. D. Eaves, J. Chem. Phys. \textbf{129}, 101104 (2008).

\bibitem{pnalbach2010njp} P. Nalbach, J. Eckel and M. Thorwart, New J. Phys. \textbf{12}, 065043 (2010).
\bibitem{cklee2015jcp} C. K. Lee, J. M. Moix, and J. S. Cao, J. Chem. Phys. \textbf{142}, 164103 (2015).
\bibitem{dzxu2016njp} D. Z. Xu, C. Wang, Y. Zhao, and J. S. Cao, New. J. Phys. \textbf{18}, 023003 (2016).

\bibitem{rjsilbey1984jcp} R. J. Silbey and T. Harris, J. Chem. Phys. \textbf{80}, 2615 (1984).

\bibitem{transitionprojector} In the eigenbasis $\{|\pm{\rangle}\}$
with $\hat{H}_s|\pm{\rangle}={\pm}\Lambda/2|\pm{\rangle}$,
the transition projectors are given by
$\hat{P}_e(\Lambda)=\cos\theta|+{\rangle}{\langle}-|$,
$\hat{P}_e(0)=\sin\theta(|+{\rangle}{\langle}+|-|-{\rangle}{\langle}-|)$,
$\hat{P}_e(-\Lambda)=\cos\theta|-{\rangle}{\langle}+|$,
$\hat{P}_o(\Lambda)=-i|+{\rangle}{\langle}-|$,
$\hat{P}_o(0)=0$,
and $\hat{P}_o(-\Lambda)=i|-{\rangle}{\langle}+|$, with $\theta=\tan^{-1}(\eta\Delta/\epsilon_0)$.


\bibitem{truokola2011prb} T. Ruokola and T. Ojanen, Phys. Rev. B \textbf{83}, 045417 (2011).
\bibitem{mesposito2009rmp} M. Esposito, U. HArbola and S. Mukamel, Rev. Mod. Phys. \textbf{81}, 1665 (2009).
\bibitem{mcampisi2011rmp} M. Campisi, P. Hanggi and P. Talkner, Rev. Mod. Phys. \textbf{83}, 771 (2011).
\bibitem{nasinitsyn2007epl} N. A. Sinitsyn and I. Nemenman, Europhys. Lett. \textbf{77}, 58001 (2007).

\bibitem{nasinitsynprl2007} N. A. Sinitsyn and I. Nemenman, Phys. Rev. Lett. \textbf{99}, 220408 (2007).
\bibitem{tsagawa2011pre} T. Sagawa and H. Hayakawa, Phys. Rev. E \textbf{84}, 051110 (2011).
\bibitem{tyuge2012prb} T. Yuge, T. Sagawa, A. Sugita and H. Hayakawa, Phys. Rev. B \textbf{86}, 235308 (2012).

\bibitem{jieren2012prl} J. Ren, S. Liu, and B. Li, Phys. Rev. Lett. \textbf{108}, 210603 (2012).
\bibitem{mberry1984prla} M. V. Berry, Proc. R. Soc. Lond. A \textbf{392}, 45 (1984).

\bibitem{rmarcus1956jcp} R. A. Marcus, J. Chem. Phys. \textbf{24}, 966 (1956).

\bibitem{refereeQ} We note that the Marcus theory is not only valid in the high temperature (classic) limit $k_BT>\omega_c$, but also valid in the relative ``high" temperature regime $\omega_c>k_BT>\epsilon_0$ (``high'' compared to the qubit's energy scale). At this latter regime, both the thermal fluctuation ($-\Gamma_vT_vt^2$) and quantum dissipation ($-i\Gamma_vt$) terms of the correlation function $Q_v(t)$ coexist, which forms the basis of our discussion.
Therefore, the Marcus theory can be carried out in this whole regime ``$k_BT>\epsilon_0$" in our work.
The related discussions can be also found at the place above ``Eq.(37) in Ref.~\cite{lnicolinjcp2011}", the place below ``Eq.(18) in Ref.~\cite{lnicolinprb2011}" and ``subsection 5.2 in Ref.~\cite{PCCP}".

\bibitem{PCCP} L. A. Pach\'on and P. Brumer, Phys. Chem. Chem. Phys., {\bf 14}, 10094 (2012).

\bibitem{note2} Please note here $Q_v(t)$ is different from the expression that people used in Ref.~\cite{tianchen2013prb} and  in Ref.~\cite{dvirasegal2006prb}. This is caused by the renormalization factor $\eta$ we extracted. They are consistent when we consider $\eta$ back.






%\bibitem{cklee2015jcp} C. K. Lee, J. M. Moix,and J. Cao, J. Chem. Phys. \textbf{142}, 164103 (2015).

\end{thebibliography}
\end{document}